\documentclass[12pt]{article}
\usepackage{graphicx}
\usepackage{epsfig}
\begin {document}
\begin{center}
\bf POMERON AND ODDERON CONTRIBUTIONS AT LHC ENERGIES

\vspace{.2cm}

C. Merino$^*$, C. Pajares$^*$, M.M. Ryzhinskiy$^{**}$, and Yu.M. Shabelski$^{**}$ \\

\vspace{.5cm}
$^*$ Departamento de F\'\i sica de Part\'\i culas, Facultade de F\'\i sica, \\ 
and Instituto Galego de F\'\i sica de Altas Enerx\'\i as (IGFAE), \\ 
Universidade de Santiago de Compostela, \\
E-mail: merino@fpaxp1.usc.es \\
E-mail: pajares@fpaxp1.usc.es \\

\vspace{.2cm}

$^{**}$ Petersburg Nuclear Physics Institute, \\
E-mail: m.ryzhinskiy@gsi.de \\
E-mail: shabelsk@thd.pnpi.spb.ru \\
\vskip 0.9 truecm


\vspace{1.2cm}

A b s t r a c t
\end{center}

We consider the first LHC data for $pp$ collisions in the framework of Regge 
theory. The integral cross sections and inclusive densities of secondaries
are determined by the Pomeron exchange, and we present the corresponding 
predictions for them. The first measurements of inclusive densities in 
midrapidity region are in agreement with these predictions. The possible 
contribution of Odderon (Reggeon with $\alpha_{Od}(0) \sim 1$ and negative 
signature) exchange to the differences in the inclusive spectra of particle 
and antiparticle in the central region could be significant at LHC energies. 
The first data of ALICE Collaboration are consistent with a very small 
Odderon contribution. Probably, further LHC data will definitely settle the
question of the Odderon existence.

\vskip 1.5cm


\newpage

\section{Introduction}

In Regge theory the Pomeron exchange dominates in the high energy soft 
hadron interaction. The Pomeron has vacuum quantum numbers. At LHC energies 
the contributions of all other exchanges to the total or inelastic cross 
sections becomes negligibly small that one can directly extract the Pomeron 
parameters directly from the experimental data.

The Quark-Gluon String Model (QGSM) \cite{KTM} is based on Dual Topological 
Unitarization (DTU), Regge phenomenology, and nonperturbative notions of QCD.
This model is successfully used for the description of multiparticle production 
processes in hadron-hadron \cite{KaPi,Sh,Ans,AMPS}, hadron-nucleus 
\cite{KTMS,Sh1}, and nucleus-nucleus \cite{JDDS} collisions. In particular, 
the inclusive densities of different secondaries produced in $pp$ collisions 
at $\sqrt{s} = 200$~GeV in midrapidity region were reasonably described in 
ref. \cite{AMPS}.

In the QGSM high energy interactions are considered as proceeding via the 
exchange of one or several Pomerons, and all elastic and inelastic processes
result from cutting through or between Pomerons \cite{AGK}. Inclusive spectra 
of hadrons are related to the corresponding fragmentation functions of quarks 
and diquarks, which are constructed using the Reggeon counting rules \cite{Kai}.
The quantitative predictions of the QGSM depend on several parameters which
were fixed by the comparison of the calculations with the experimental data
obtained at fixed target energies.

The first experimental data obtained at LHC allow one to test the stability of 
the QGSM predictions and of the values of the parameters. Fortunately, we see
that the model prediction are in reasonable agreement with the data so there
is no reason for the corrections in the parameter values.

The difference in the the total interaction cross sections and in the 
inclusive spectra of antiparticles and particles is governed by the 
numerically small contributions of Regge-poles with negative signature. 
A well-known such a Regge-pole is the $\omega$-reggeon with 
$\alpha_{\omega}(0) \sim 0.4-0.5$. Due to the small value of 
$\alpha_{\omega}(0)$ its contribution at LHC energies should be very small.  

The Odderon is a singularity in the complex $J$-plane with intercept 
$\alpha_{Od} \sim 1$, negative $C$-parity, and negative signature. Thus, 
its zero flavour-number exchange contribution to particle-particle and 
to antiparticle-particle interactions, e.g., to $pp$ and to $\bar{p}p$ 
total cross sections, or to the inclusive production of baryons and of 
antibaryons in $pp$ collisions has opposite signs. In QCD the Odderon 
singularity is connected~\cite{BLV} to the colour-singlet exchange of three 
reggeized gluons in $t$-channel. The theoretical and experimental status of 
Odderon has been recently discussed in refs.~\cite{Nic,Ewe}. The possibility 
to detect Odderon effects has also been investigated in other domains, as the 
$\pi p \to \rho N$ reaction~\cite{Cont} and charm photoproduction~\cite{BMR}.

The Odderon coupling should be very small with respect to the Pomeron 
coupling. However, several experimental facts favouring the presence of the 
Odderon contribution exist, e.g., the energy behaviour of the difference of 
total $\bar{p}p$ and $pp$ cross sections \cite{MRS,MRS1}, and the 
difference in the $d\sigma/dt$ behaviour of elastic $pp$ and $\bar{p}p$ 
scattering at $\sqrt{s} =$ 52.8 GeV and $\vert t \vert = 1. - 1.5$ GeV$^2$ 
presented in references~\cite{Nic,Bre}. The behaviour of $pp$ and $\bar{p}p$ 
elastic scattering and total cross sections at ISR and SPS energies was 
analyzed in~\cite{JSS}. 

The differences in the yields of baryons and antibaryons produced in the 
central (midrapidity) region  of high energy $pp$ interactions 
\cite{AMPS,MRS,MRS1,ACKS,BS,AMS,Olga,SJ3} can also be significant in this 
respect. The question of whether the Odderon exchange is needed for 
explaining these experimental facts, or they can be described by the usual 
exchange of a reggeized quark-antiquark pair with 
$\alpha_{\omega}(t) = \alpha_{\omega}(0) + \alpha_{\omega}'t$, it 
 is a fundamental one. 

In this paper we present the description of the first LHC data in the framework 
of QGSM, as well as some predictions for the Pomeron and Odderon effects at 
LHC energies

\section{Cross sections at LHC energies}

Let us start with the analysis of high energy elastic particle and antiparticle 
scattering on the proton target. Here, the simplest contribution is the one 
Regge-pole $R$ exchange corresponding to the scattering amplitude
\begin{equation}
A(s,t) = g_1(t)\cdot g_2(t)\cdot \left(\frac{s}{s_0}\right)^{\alpha_R(t) - 1}ç
\cdot \eta(\Theta) \;,
\end{equation}
where $g_1(t)$ and $g_2(t)$ are the couplings of a Reggeon to the beam and 
target hadrons, $\alpha_R(t)$ is the $R$-Reggeon trajectory, and 
$\eta(\Theta)$ is the signature factor which determines the complex structure 
of the scattering amplitude ($\Theta$ equal to +1 and to $-1$ for reggeon with 
positive and negative signature, respectively):
\begin{equation}
\eta(\Theta) = \left\{ \begin{array}{ll} i - \tan^{-1}(\frac{\pi \alpha_R}2) 
& \Theta = +1 \\
i + \tan({\frac{\pi \alpha_R}2}) & \Theta = -1 \;, \end{array} \right.
\end{equation}
so the amplitude $A(s,t=0)$ becomes purely imaginary for positive signature 
and purely real for negative signature when $\alpha_R \to 1$. 

The interaction of a particle or of an antiparticle with a proton target
is the same for the Reggeon exchange with positive signature, but in the case 
of negative signature the two contributions have opposite signs, as it is 
shown in Fig.~1.
\begin{figure}[htb]
\centering
\vskip -.2cm
\includegraphics[width=.7\hsize]{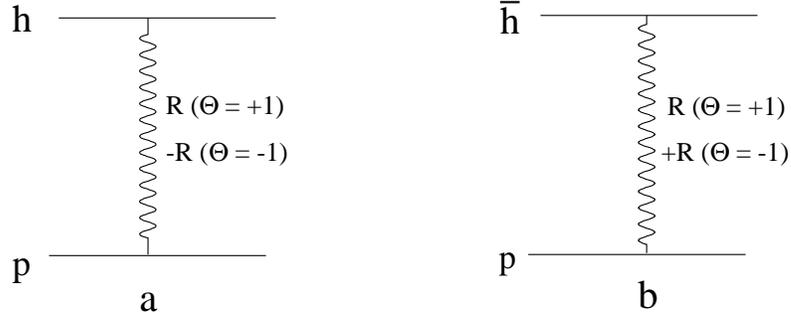}
\vskip -.5cm
\caption{\footnotesize
Diagram corresponding to the Reggeon-pole exchange in particle $h$ (a) (its 
antiparticle $\bar{h}$ (b)) interactions with a proton target. The positive 
signature ($\Theta = +1$) exchange contributions are the same, while the 
negative signature ($\Theta = -1$) exchange contributions have opposite signs.}
\end{figure}

The corresponding pole trajectory is given by
\begin{equation}
\alpha_R (t) = \alpha_R(0) + \alpha'_R \, t \;,
\end{equation}
where $\alpha_R(0)$ (intercept) and $\alpha'_R$ (slope) are some numbers.

In the case of Pomeron trajectory with $\alpha_P(0) > 1$ the correct 
asymptotic behavior $\sigma_{tot} \sim \ln^2s$ \cite{Volk,Kop}, compatible 
with the Froissart bound can only be obtained by taking into account the 
multipomeron cuts.

\begin{figure}[htb]
\centering
\mbox{\psfig{file=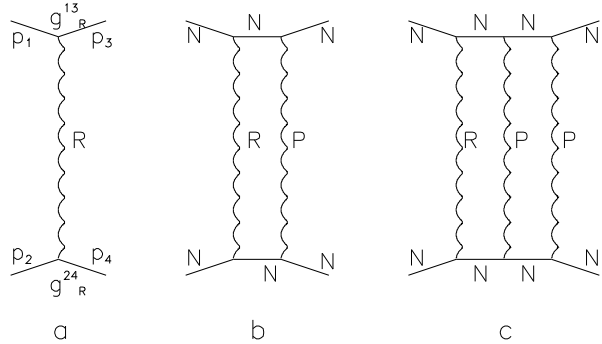,width=0.65\textwidth}} \\
\caption{\footnotesize
Regge-pole theory diagrams:  (a) single $R$-pole exchange in
the binary process $1+2 \to 3+4$, double $R P$ (b) and triple $R P P$
(c) exchanges in elastic $N N$ scattering.}
\end{figure}

Indeed, for the Pomeron trajectory
\begin{equation}
\alpha_P(t) = 1 + \Delta + \alpha'_P \, t\;,\,\, \Delta > 0 \;,
\end{equation}
the one-Pomeron contribution to $\sigma^{tot}_{hN}$ equals
\begin{equation}
\sigma_P = 8\pi \gamma e^{\Delta \cdot \xi}, \; {\rm with}\; \xi = \ln s/s_0 \;,
\end{equation}
where $\gamma = g_1(0) \cdot g_2(0)$ is the Pomeron coupling, 
$s_0 \simeq 1$ GeV$^2$, and $\sigma_P$ rises with energy as $s^{\Delta}$. 
To obey the $s$-channel unitarity, and the Froissart bound in particular, 
this contribution should be screened by the multipomeron discontinuities. 
A simple quasi-eikonal treatment \cite{Kar3} yields to
\begin{equation}
\sigma^{tot}_{hN} = \sigma_P f(z/2) \;,\,\, \; \sigma^{el}_{hN} =
\frac{\sigma_P}{C} [f(z/2) - f(z)] \;,
\end{equation}
\begin{equation}
f(z)  =  \sum^{\infty}_{k=1} \frac1{k \cdot k !} (-z)^{k-1} =
\frac1z \int ^z_0 \frac{dx}x (1-e^{-x}) \;, 
\end{equation}
\begin{equation}
z = \frac{2C\gamma}{\lambda} e^{\Delta \xi} \;,\,\, \lambda = R^2 + \alpha'_P
\xi \;.
\end{equation}

The numerical values of the Pomeron parameters were taken \cite{Volk,Sh}
to be :
\begin{equation}
\Delta = 0.139 , \;\;  \alpha'_P = 0.21 {\rm GeV}^{-2}, \;\;
\gamma = 1.77 {\rm GeV}^{-2}, \;\; R^2 = 3.18 {\rm GeV}^{-2}, \;\; C = 1.5. 
\end{equation}
Here, $R^2$ is the radius of the Pomeron and $C$ is the quasi-eikonal
enhancement coefficient (see \cite{Kar1}).

The predictions of Regge theory with obtained these values of the parameters 
(a small comtribution from non-Pomeron exchange with parameters taken from
\cite{Volk} is accounted for) are presented in Table 1.

\begin{center}


\vskip 5pt
\begin{tabular}{|c||r|r|r|} \hline
$\sqrt{s}$  & $\sigma^{tot}$ & $\sigma^{el}$ & $\sigma^{inel}$ \\ \hline
900 GeV & 67.4 & 13.2 & 54.2  \\
7 TeV   & 94.5 & 21.1 & 73.4  \\
14 TeV & 105.7 & 24.2 & 81.5  \\

\hline
\end{tabular}

\end{center}

\noindent
Table 1. The Regge theory predictions for total, total elastic and total 
inelastic cross sections (in mb) in $pp$ collisions at LHC energies. 

At asymptotically high energies ($z \gg 1$) we obtain
\begin{equation}
\sigma^{tot}_{hN} = \frac{8\pi \alpha'_P \Delta}{C} \xi^2 \;,\,\, \;
\sigma^{el}_{hN} = \frac{4\pi \alpha'_P \Delta}{C^2} \xi^2 \;,
\end{equation}
according to the Froissart limit \cite{Froi}.

However, in the complete Reggeon diagram technique \cite{Grib} not only
Regge-poles and cuts, but more complicated diagrams (e.g. the  so-called
enhanced diagrams of the type of Fig.~3) should be taken into account. In 
the numerical calculations of such diagrams some new uncertainties appear, 
because the vertices of the coupling of $n$ and $m$ Reggeons 
(see Fig.~3c) are unknown, so some model estimations are needed. 

\begin{figure}[htb]
\centering
\vspace {-.5cm}
\mbox{\psfig{file=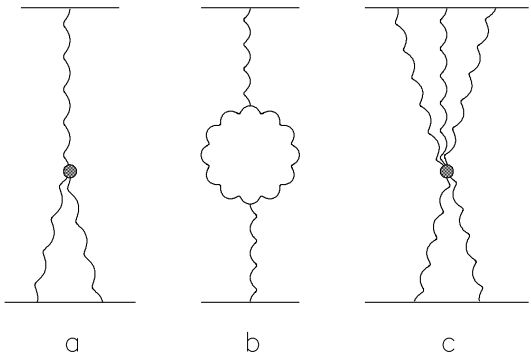,width=0.6\textwidth}} \\
\caption{\footnotesize
Examples of enhanced Reggeon diagrams.}
\end{figure}

The common feature of the calculations in which enhanced diagrams are 
included results in the additional increase of the Pomeron intercept 
$\alpha_P(0) = 1 + \Delta$, needed for the description of the experimental 
data. For example, a value $\Delta = 0.21$ was obtained in \cite{KPKar}.
In models with strong interactions among the produced strings also a larger
$\Delta$ is obtained. Thus, in string percolation $\Delta = 2/7$
\cite{BDP}. Also in color glass condensate $\Delta$ is large,
$\Delta = 0.28$ \cite{ASW}.
On the other hand, the QCD solution corresponding to a bare Pomeron is known to the leading 
log accuracy \cite{FKL,BL,GLR,Lip}:
\begin{equation}
\Delta = N_c \frac{\alpha_s}{\pi} 4\ln2 \\;,
\end{equation}
where $N_c$ is the number of colors. However, the next-to leading
corrections to these solutions are very large \cite{FL,ABB}. The
situation with the Pomeron intercept is not clear \cite{GSAl} up to now.
The value of the Pomeron slope is also discussed. Phenomenologically,in 
many papers it is chosen to be $\alpha'_P \sim 0.25$ GeV$^{-2}$. 
A more theoretical (QCD) point of view with
$\alpha'_P \to 0$ at $s \to \infty$ is presented in \cite{LRa}.

The numerical calculations which account for enhanced diagrams 
\cite{KMR,GL,Ost} lead to the values of $\sigma^{inel}$ of the same order 
($\pm$ 10\% at $\sqrt{s} = 14$ TeV) as those presented in Table.~1.

\section{Inclusive spectra in midrapidity region}

The inclusive cross section for the production of a secondary $h$ in high 
energy $pp$ collisions in the central region is determined by the Regge-pole 
diagrams shown in Fig.~4~\cite{AKM}. The diagram with only Pomeron exchange 
(Fig.~4a) is the leading one, while the diagrams with one secondary Reggeon 
$R$ (Figs.~4b and 4c) correspond to corrections which disappear with the 
increase of the initial energy.

\begin{figure}[htb]
\centering
\vskip -2.5cm
\includegraphics[width=.55\hsize]{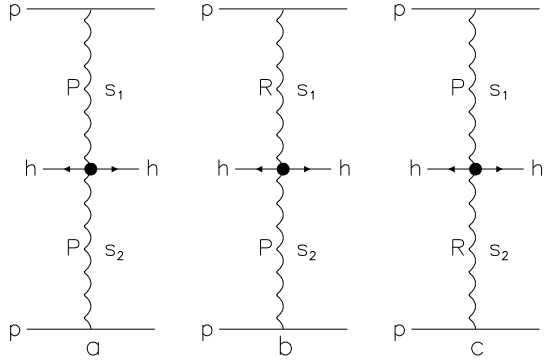}
\caption{\footnotesize 
Regge-pole diagrams for the inclusive production of a secondary hadron $h$ 
in the central region.}
\end{figure}

The inclusive production cross section of hadron $h$ with transverse momentum 
$p_T$ corresponding to the diagram shown in Fig.~4b has the following expression:
\begin{equation}
F(p_T,s_1,s_2,s) = \frac{1}{\pi^2 s} g^{pp}_{R}\cdot g^{pp}_{P}\cdot 
g^{hh}_{RP}(p_T) \cdot \left(\frac{s_1}{s_0} \right)^{\alpha_{R}(0)}\cdot \left
(\frac{s_2}{s_0}\right)^{\alpha_{P}(0)} \;,
\end{equation}
where
\begin{eqnarray}
s_1 & = & (p_a + p_h)^2 = m_T\cdot s^{1/2}\cdot e^{-y} \\ \nonumber
s_2 & = & (p_b + p_h)^2 = m_T\cdot s^{1/2}\cdot e^{y} \;,
\end{eqnarray}
with $s_1\cdot s_2 = m^2_T\cdot s$ \cite{Kar}, and the rapidity $y$ 
defined in the center-of-mass frame. 

At very high energies, only the one-Pomeron exchange diagram in Fig.~4a 
contributes to the inclusive density in the central region 
$y \sim 0$ (AGK theorem \cite{AGK}). This leads to

\begin{equation}
\frac{d\sigma}{dy} \sim (\frac{s}{s_0})^{\alpha_P(0) - 1} = 
(\frac{s}{s_0})^{\Delta_P}  \;.
\end{equation}

Now one can estimate~\cite{LUCh} the intercept of the supercritical Pomeron by considering  
\begin{equation}
\frac{d\sigma}{dy} = \sigma^{pp}_{inel} \frac{dn}{dy} \;,
\end{equation}
and by defining
\begin{equation}
\Delta_P =  \Delta_{\sigma} + \Delta_n \;,
\end{equation}
where $\Delta_{\sigma}$ comes from the energy dependence of $\sigma^{pp}_{inel}$
and $\Delta_n$ corresponds to the energy dependence of $dn/dy$.

The analysis of the $dn/dy$ energy behaviour in the energy interval 
$\sqrt{s} = 15 - 900$ GeV was provided in ref. \cite{UA5}, and it corresponds
to a value: 
\begin{equation}
\Delta_n = 0.105 \pm 0.006 \;.
\end{equation}

From recent LHC (ALICE Collaboration) data $\sqrt{s} = 900$ GeV -- 7 TeV,
one obtains 
\begin{equation}
\Delta_n = 0.110 \pm 0.008 \:,
\end{equation}
and showing a very stable behaviour. 

For the value of $\Delta_{\sigma}$ in the last energy interval Regge theory 
predicts  
\begin{equation}
\Delta_{\sigma} = 0.07 \;,
\end{equation}
so for $\Delta_P$ we obtain 
\begin{equation}
\Delta_P = 0.18 \;.
\end{equation}

The only problem is that the values of $dn_{ch}/d\eta$ in ref. \cite{ALICE} 
were obtained under the condition that in the considered kinematical window as 
a minimum one charged particle should exist. This condition increase the value 
of $dn_{ch}/d\eta$ at $\sqrt{s} = 900$ GeV more significantly than at 
$\sqrt{s} = 7$ TeV, so the value of $\Delta$ presented in Eq.~(15) should 
be slightly increased. 

In the case of only Pomeron exchange, Fig.~4a, the yields of particle and 
antiparticle in the central region are equal. The difference between them 
comes from the first correction to the Pomeron diagram. This correction is
shown in Figs.~4b, 4c, where $R$ is the effective sum of all amplitudes 
with negative signature ($\Theta = -1$ in Eq.~(2), so its contribution to 
the inclusive spectra of secondary protons and antiprotons has the opposite 
sign. In the midrapidity region, i.e. at $y \sim 0$, the ratios of 
$\bar{p}$ and $p$ (or any other antibaryon and baryon) yields integrated over 
$p_T$($\langle m_T \rangle \simeq 1$ GeV) can be written as
\begin{equation}
\frac{\bar{p}}p = \frac{1 - r_-(s)}{1 + r_-(s)}  \;,
\end{equation}
where $r_-(s)$ is the ratio of the negative signature ($R$) to the positive 
signature ($P$) contributions \cite{MRS,MRS1}:
\begin{equation}
r_-(s) = c_1\cdot \left(\frac{s}{s_0}\right)^{(\alpha_{R}(0) - \alpha_{P}(0))/2} \;.
\end{equation}
Here $c_1$ is a normalization constant and the physically important quantity 
is the difference of intercepts $(\alpha_{R}(0) - \alpha_{P}(0))$, which can be 
determined from the comparison with the experimental data.
.

\section{Inclusive spectra of secondary hadrons \newline in the
Quark-Gluon String Model}

The Quark-Gluon String Model (QGSM) \cite{KTM,KaPi,Sh} allows us to make 
quantitative predictions of different features of multiparticle production, 
in particular, the  inclusive densities of different secondaries both in the 
central and in beam fragmentation regions. In QGSM high energy hadron-nucleon 
collisions are considered as taking place via the exchange of one or several 
Pomerons, all elastic and inelastic processes resulting from cutting through 
or between Pomerons~\cite{AGK}. 

Each Pomeron corresponds to a cylindrical diagram (see Fig.~5a), and thus, when 
cutting one Pomeron, two showers of secondaries are produced as it is shown in 
Fig.~5b. The inclusive spectrum of a secondary hadron $h$ is then determined 
by the convolution of the diquark, valence quark, and sea quark distributions 
$u(x,n)$ in the incident particles, with the fragmentation functions $G^h(z)$ 
of quarks and diquarks into the secondary hadron $h$. These distributions, as 
well as the fragmentation functions are constructed using the Reggeon counting 
rules \cite{Kai}. Both the diquark and the quark distribution functions depend 
on the number $n$ of cut Pomerons in the considered diagram.

\begin{figure}[htb]
\centering
\includegraphics[width=.6\hsize]{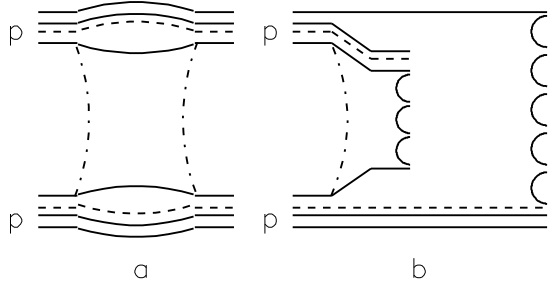}
\caption{\footnotesize
(a) Cylindrical diagram corresponding to the one--Pomeron exchange 
contribution to elastic $pp$ scattering, and (b) the cut of this diagram which 
determines the contribution to the inelastic $pp$ cross section (b). Quarks 
are shown by solid curves and string junction by dashed curves.}
\end{figure}

For a nucleon target, the inclusive rapidity (y) or Feynman-$x$ ($x_F$) 
spectrum of a secondary hadron $h$ has the form~\cite{KTM}:
\begin{equation}
\frac{dn}{dy}\ = \
\frac{x_E}{\sigma_{inel}}\cdot \frac{d\sigma}{dx_F}\ = \frac{dn}{dy} =
\sum_{n=1}^\infty w_n\cdot\phi_n^h (x) \ ,
\end{equation}
where the functions $\phi_{n}^{h}(x)$ determine the contribution of the 
diagram with $n$ cut Pomerons and $w_n$ is the relative weight of this 
diagram. Here we neglect the contribution of diffraction dissociation 
processes which is very small in the midrapidity region.

For $pp$ collisions
\begin{eqnarray}
\phi_{pp}^h(x) & = & f_{qq}^{h}(x_+,n)\cdot f_q^h(x_-,n) +
f_q^h(x_+,n)\cdot f_{qq}^h(x_-,n) + \nonumber\\
& + &  2(n-1)f_s^h(x_+,n)\cdot f_s^h(x_-,n)\ ,
\end{eqnarray}

\begin{equation}
x_{\pm} = \frac12\left[\sqrt{4m_T^2/s+x^2}\ \pm x\right] ,
\end{equation}
where $f_{qq}$, $f_q$, and $f_s$ correspond to the contributions of diquarks, 
valence quarks, and sea quarks, respectively.

These functions are determined by the convolution of the diquark and quark 
distributions with the fragmentation functions, e.g. for the quark one can 
write:
\begin{equation}
f_q^h(x_+,n)\ =\ \int\limits_{x_+}^1u_q(x_1,n)\cdot G_q^h(x_+/x_1) dx_1\ .
\end{equation}
The diquark and quark distributions, which are normalized to unity, as well 
as the fragmentation functions, are determined by the corresponding Regge 
intercepts 
\cite{Kai}.

At very high energies both $x_+$ and $x_-$ are negligibly small in the 
midrapidity region, and so all fragmentation functions, which are usually 
written \cite{Kai} as $G^h_q(z) = a_h (1-z)^{\beta}$, become constants and 
equal for a particle and its antiparticle: 
\begin{equation}
G_q^h(x_+/x_1) = a_h \ . 
\end{equation}
This leads, in agreement with \cite{AKM} and with Eq.~(14), to
\begin{equation}
\frac{dn}{dy}\ = \ g_h \cdot (s/s_0)^{\alpha_P(0) - 1}
\sim a^2_h \cdot (s/s_0)^{\alpha_P(0) - 1} \,,
\end{equation}
that corresponds to the only one-Pomeron exchange diagram in Fig.~4a, i.e. to 
the only diagram contributing to the inclusive density in the central region 
(AGK theorem \cite{AGK}) at asymptotically high energy. The values of the 
Pomeron parameters presented in Eq.~(9) are used in the QGSM numerical 
calculations. 

The QGSM predictions for the initial energy dependence of the inclusive 
densities $dn/d\eta \vert_{\eta = 0}$ for all charged secondaries produced in 
high energy $pp$ collisions are presented in Fig.~6.

\begin{figure}[htb]
\centering
\includegraphics[width=.75\hsize]{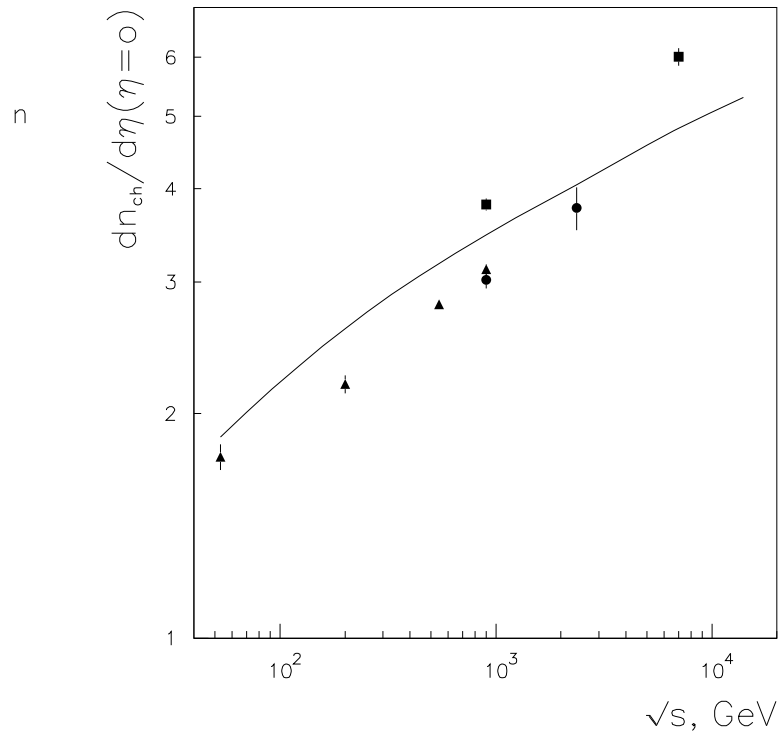}
\caption{\footnotesize 
The QGSM predictions for the inclusive densities in the midrapidity region
for all charged secondaries as a function of the initial energy. Old data of 
ISR and SpS for all inelastic $pp$ ($\bar{p}p$) collisions \cite{UA5} are shown 
by triangles, ALICE data \cite{ALIC} for all inelastic collisions by points, 
and ALICE data \cite{ALICE} for events with INEL$>0$ by squares (see full text).}
\end{figure}

The theoretical curve slightly overestimates the data for all inelastic 
collisions. This small disagreement rests inside the model accuracy. At the
same time, the curve lies below the points of ALICE Collaboration shown by 
squares \cite{ALICE}, which were obtained for the events INEL$> 0$, that is 
for events in which as minimum one charged particle should be detected in the 
kinematical window $\vert \eta \vert > 1$ (see \cite{ALICE}).  Thus the 
events without any charged particle in this kinematical window are 
ignored, what evidently increases the inclusive density.

It is necessary to note that the calculated values of $dn/d\eta (\eta = 0)$ 
depend on the averaged transverse momenta of secondaries. The values 
$\langle p_T \rangle _{\pi}$ = 0.35 GeV/c, $\langle p_T \rangle _K$ = 0.52 
GeV/c, and  $\langle p_T \rangle _p$ = 0.68 GeV/c \cite{abe} were used in the 
Fig.~6 for energies $\sqrt{s} \geq 900$ GeV. The values of $dn/d\eta$ would 
increase with energy slightly faster if the averaged transverse momenta 
of secondaries would increase.

The QGSM predictions for the $dn/d\eta$ distributions of all charged 
secondaries produced in inelastic $pp$ and $\bar{p}p$ collisions at 
different energies are shown in Fig.~7. The experimental data are taken 
from \cite{Kaid}.

\begin{figure}[htb]
\centering
\includegraphics[width=.75\hsize]{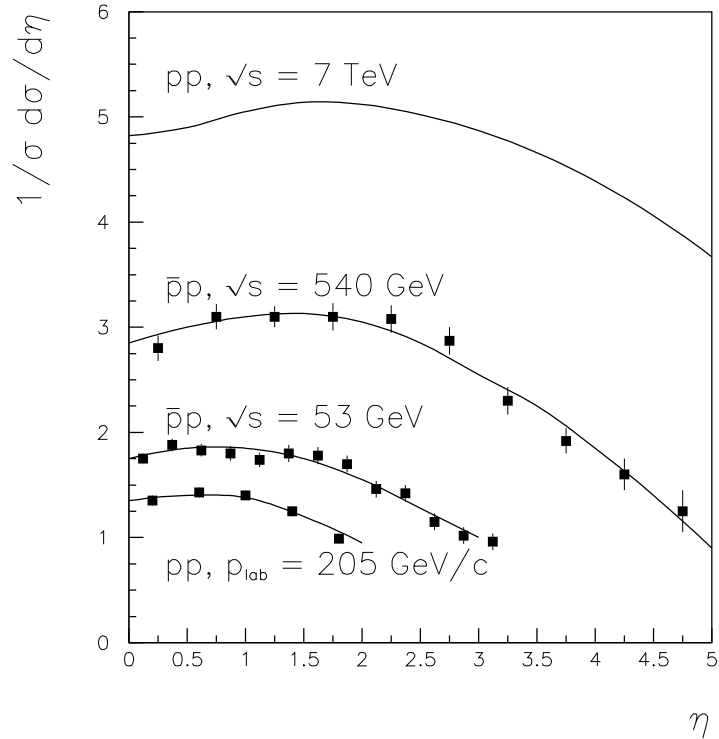}
\caption{\footnotesize 
The QGSM predictions for the pseudorapidity distributions of all charged 
secondaries produced in inelastic $pp$ and $\bar{p}p$ collisions at 
different energies.}
\end{figure}

The QGSM allows one to calculate the inclusive spectra of different 
secondaries. In Fig.~8 we compare the QGSM predictions for the integral 
multiplicities of charged kaons which are compiled in ref. \cite{NA49}. 
The agreement with the existing data for $pp$ collisions in the energy 
interval $\sqrt{s} = 17-200$ GeV is good. 

\begin{figure}[htb]
\centering
\vskip -.5cm
\includegraphics[width=.75\hsize]{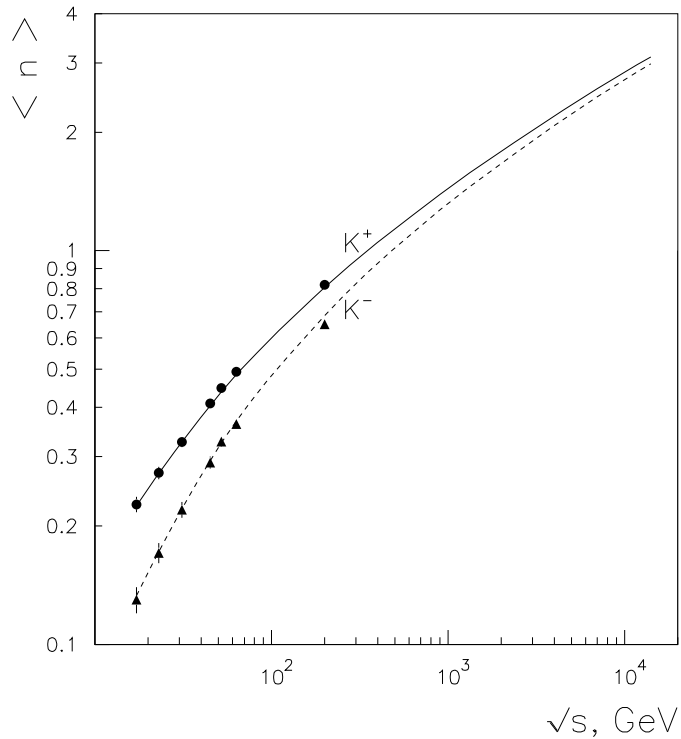}
\caption{\footnotesize 
The QGSM predictions for the integral multiplicities of $K^+$ and $K^-$ 
as a functions of the initial energy.}
\end{figure}

In the kaon sector the most interesting case is that of the $K^0_s$-mesons, 
which can be used for measurement of CP-violation, etc. In Fig.~9 we present 
the experimental data on the midrapidity inclusive densities of $K^0_s$-mesons
produced in $pp$ and $\bar{p}p$ collisions at different energies ref. 
\cite{NA49}, together with the results of QGSM calculations. 

\begin{figure}[htb]
\centering
\includegraphics[width=.75\hsize]{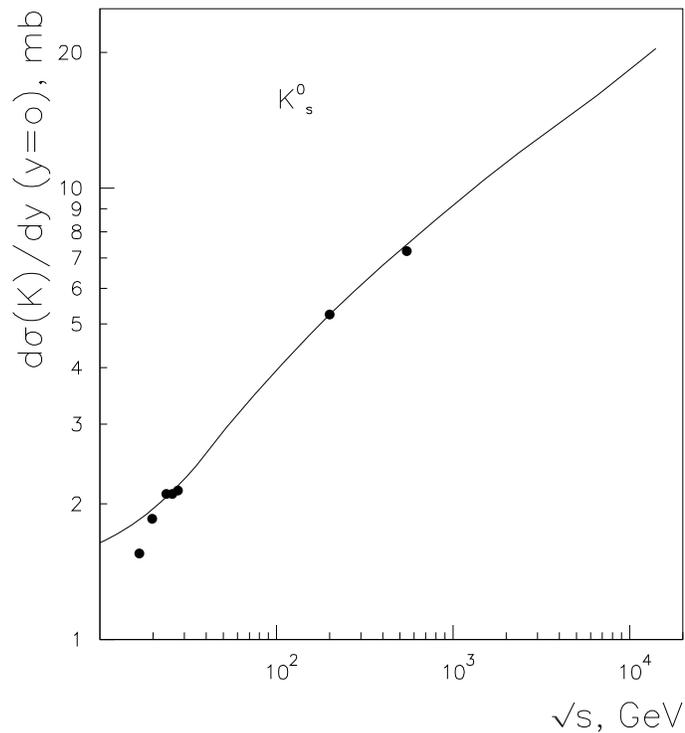}
\caption{\footnotesize
The QGSM predictions for the inclusive densities in the midrapidity 
region for $K^0_s$-mesons produced in $pp$ and $\bar{p}p$ collisions
as a function of the initial energy.}
\end{figure}

In Fig.~10 we compare the QGSM predictions for the  rapidity distributions
of $K^0_s$ produced in $pp$ collisions at $\sqrt{s} = 900$ GeV (solid 
curve) with the experimental data of LHCb Collaboration \cite{LHCb},
together with RHIC (PHENIX and STAR Collaboration) data \cite{RHIC},
as well as the QGSM predictions at $\sqrt{s} = 200$ GeV (dashed curve).

\begin{figure}[htb]
\centering
\includegraphics[width=.55\hsize]{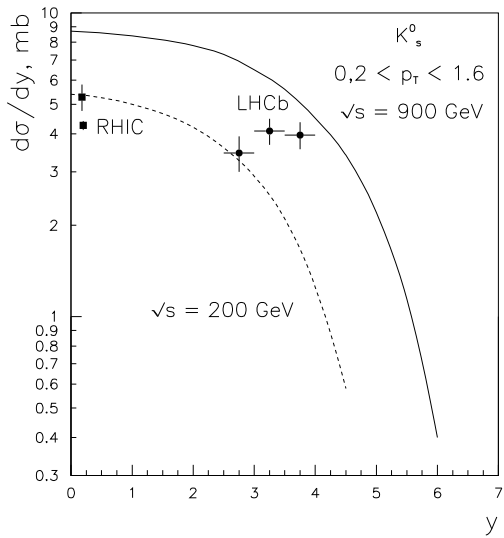}
\vskip -.3cm
\caption{\footnotesize 
The QGSM predictions for the inclusive cross sections of secondary 
$K^0_s$ produced in $pp$ collisions at $\sqrt{s} = 900$ GeV (solid
curve) and at $\sqrt{s} = 200$ GeV (dashed curve).}
\end{figure}

The theoretical curve for RHIC energy is in agreement with the experimental
data \cite{RHIC}. The curve for LHCb clearly overestimates the inclusive 
cross section of $K^0_s$ production, but it is necessary to keep in mind 
that the curve corresponds to the $d\sigma/dy$ values integrated over 
transverse momenta, while the data \cite{LHCb} were obtained in the 
region $0.2 < p_T <1.6$ GeV/c, so they will be increased after accounting 
for the contributions of very low and high $p_T$.

\section{Baryon/antibaryon asymmetry in QGSM}

The difference in the total cross section of high energy particle and 
antiparticle scattering on the proton target is (see Fig.~1)
\begin{equation}
\Delta \sigma^{tot}_{hp} = \sum_{R(\Theta=-1)} 2\cdot Im\,A(s,t=0) =
\sum_{R(\Theta=-1)} 2\cdot g_1(0)\cdot g_2(0)\cdot  
\left(\frac{s}{s_0}\right)^{\alpha_R(0) - 1}
\cdot Im\,\eta(\Theta=-1) \;.
\end{equation}
The experimental data for the differences of $\bar{p}p$ and $pp$ total cross 
sections at $\sqrt{s} > 8$ GeV are presented in Fig.~11. Here we use the data 
compiled in ref.~\cite{CERN} by presenting at every energy the experimental 
points for $pp$ and $\bar{p}p$ by the same experimental group and with the 
smallest error bars. At ISR energies (last three points in Fig.~11) we 
present the data in ref.~\cite{Car} as published in their most recent 
version of ref.~\cite{CERN}.

\begin{figure}[htb]
\centering
\vskip .2cm
\includegraphics[width=.5\hsize]{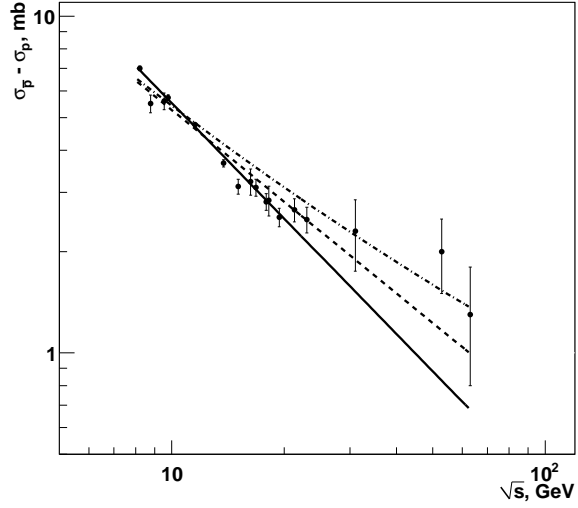}
\caption{\footnotesize 
Experimental differences of $\bar{p}p$ and $pp$ total cross sections at 
$\sqrt{s} > 8$ GeV, together with their one-Reggeon fit (solid curve), 
fit of \cite{DL} (dashed curve), and fit by the sum of $\omega$-Reggeon
and Odderon contribution (dash-dotted curve).}
\end{figure}

From the results of this fit one can see that the usual one-power fit 
\cite{ShSh,DL} of $\Delta \sigma^{tot}_{pp}$ by only $\omega$-Reggeon is 
not good enough, and one additional Odderon contribution with 
$\alpha_{Odd}(0) \sim 0.9$ is in agreement with the experimental data. The 
contributions of Odderon and $\omega$-reggeon to the differences in $\bar{p}p$ 
and $pp$ total cross sections would be approximately equal at 
$\sqrt{s} \sim 25-30$ GeV. In any case, a more detailed analysis is needed, 
especially concerning the experimental error bars for the differences in $pp$ 
and $\bar{p}p$ cross sections.

In the string models, baryons are considered as configurations consisting of 
three connected strings (related to three valence quarks) called  string 
junction (SJ) \cite{Artru,IOT,RV,Khar}, as it is shown in Fig.~12.

\begin{figure}[htb]
\centering
\vskip -2.cm
\includegraphics[width=.5\hsize]{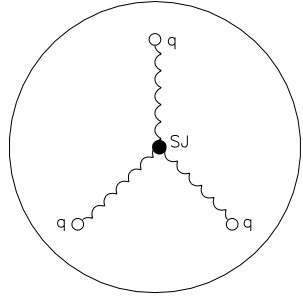}
\vskip -1.2cm
\caption{\footnotesize
The composite structure of a baryon in string models. Quarks are shown by open 
points and SJ by black point.}
\end{figure}

The colour part of a baryon wave function reads as follows~\cite{Artru,RV}:
\begin{eqnarray}
&&B\ =\ \psi_i(x_1)\cdot\psi_j(x_2)\cdot\psi_k(x_3)\cdot J^{ijk}(x_1, x_2, x_3, x) 
\,, \\
&& J^{ijk}(x_1, x_2, x_3, x) =\ \Phi^i_{i'}(x_1,x)\cdot\Phi_{j'}^j(x_2,x)\cdot
\Phi^k_{k'}(x_3,x)\cdot\epsilon^{i'j'k'} \,, \\
&& \Phi_i^{i'}(x_1,x) = \left[ T\cdot\exp \left(g\cdot\int\limits_{P(x_1,x)}
A_{\mu}(z) dz^{\mu}\right) \right]_i^{i'} \,,
\end{eqnarray}
where $x_1, x_2, x_3$, and $x$ are the coordinates of valence quarks and SJ, 
respectively, and $P(x_1,x)$ represents a path from $x_1$ to $x$ which looks 
like an open string with ends at $x_1$ and $x$. Such a baryon structure is supported
by lattice calculations \cite{latt}. 

This picture leads to some general phenomenological predictions. In
particular, it opens room for exotic states, such as the multiquark bound 
states, 4-quark mesons, and pentaquarks \cite{RV,DPP1,RSh}. In the case of 
inclusive reactions the baryon number transfer to large rapidity distances in 
hadron-nucleon and in hadron-nucleus reactions can be explained \cite{ACKS} 
by SJ diffusion.

The production of a baryon-antibaryon pair in the central region usually occurs
via $SJ$-$\overline{SJ}$ pair production (according to Eqs.~(30), (31), SJ has 
upper color indices, whereas antiSJ ($\overline{SJ}$) has lower indices) which 
then combines with sea quarks and sea antiquarks into, respectively, $B\bar{B}$ 
pair \cite{RV,VGW}, as it is shown in Fig.~13. 

\begin{figure}[htb]
\centering
\vskip -3.5cm
\hspace{4.5cm}
\includegraphics[width=.6\hsize]{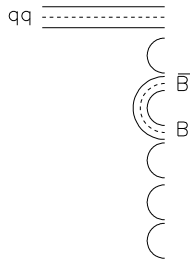}
\vskip -1.8cm
\caption{\footnotesize
Diagram corresponding to the diquark fragmentation function for the 
production of a central $\bar{B}B$ pair. Quarks are shown by solid curves and SJ by
dashed curves.}
\end{figure}

In processes with incident baryons, say, in $pp$ collisions there
exists another possibility to produce a secondary baryon  in the central
region. This second possibility leads to the diffusion in rapidity space of
the two SJ existing in the initial state that leads to significant differences 
in the yields of baryons and antibaryons in the midrapidity region even at 
rather high energies~\cite{ACKS,AMS}. Probably, the most important 
experimental fact in favour for this process is the rather large asymmetry in 
$\Omega$ and $\bar{\Omega}$ baryon production in high energy $\pi^-p$ 
interactions \cite{ait}. 

The quantitative theoretical description of the baryon number transfer 
via SJ mechanism was suggested in the 90's and used to predict~\cite{KP1} the 
$p/\bar{p}$ asymmetry at HERA energies, which was experimentally observed
\cite{H1}. It was also noted in ref.~\cite{Bopp} that the $p/\bar{p}$ 
asymmetry measured at HERA can be obtained by simple extrapolation of ISR data.
The quantitative description of the baryon number transfer due to SJ diffusion 
in rapidity space was firstly obtained in \cite{ACKS} and following papers 
\cite{AMPS,BS,AMS,Olga,SJ3,MRS,MRS1}. 

In the QGSM the differences in the spectra of secondary baryons and 
antibaryons produced in the central region appear for processes which present 
SJ diffusion in rapidity space. These differences only vanish rather slowly 
when the energy increases. 

To obtain the net baryon charge we consider according to ref.~\cite{ACKS}, we 
consider three different possibilities. The first one is the fragmentation of 
the diquark giving rise to a leading baryon (Fig.~14a). A second possibility 
is to produce a leading meson in the first break-up of the string and one 
baryon in a subsequent break-up~\cite{Kai,22r} (Fig.~14b). In these two first 
cases the baryon number transfer is possible only for short distances in 
rapidity. In the third case, shown in Fig.~14c, both initial valence quarks 
in the diquark recombine with sea antiquarks into mesons $M$, while a secondary 
baryon is formed by the SJ together with three sea quarks. 

\begin{figure}[htb]
\centering
\includegraphics[width=.6\hsize]{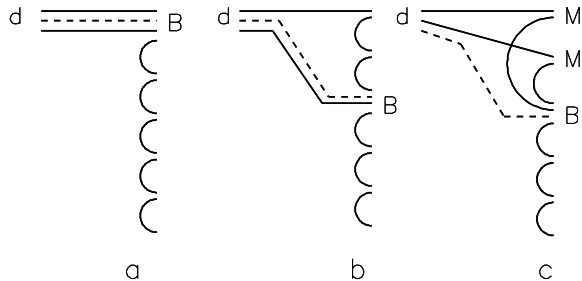}
\caption{\footnotesize
QGSM diagrams describing secondary baryon $B$ production by diquark $d$: 
initial SJ together with two valence quarks and one sea quark (a), initial SJ 
together with one valence quark and two sea quarks (b), and initial SJ 
together with three sea quarks (c).}
\end{figure}

The fragmentation functions for the secondary baryon $B$ production 
corresponding to the three processes shown in Fig.~14 can be written 
as follows (see~\cite{ACKS} for more details):
\begin{eqnarray}
G^B_{qq}(z) &=& a_N\cdot v^B_{qq} \cdot z^{2.5} \;, \\
G^B_{qs}(z) &=& a_N\cdot v^B_{qs} \cdot z^2\cdot (1-z) \;, \\
G^B_{ss}(z) &=& a_N\cdot\varepsilon\cdot v^B_{ss} \cdot z^{1 - \alpha_{SJ}}\cdot 
(1-z)^2  \;,
\end{eqnarray}
for Figs.~14a, 14b, and 14c, respectively, and where $a_N$ is the normalization 
parameter, and $v^B_{qq}$, $v^B_{qs}$, $v^B_{ss}$ are the relative probabilities 
for different baryons production that can be found by simple quark 
combinatorics \cite{AnSh,CS}. 

The fraction $z$ of the incident baryon energy carried by the secondary baryon 
decreases from Fig.~14a to Fig.~14c, whereas the mean rapidity gap between the 
incident and secondary baryon increases. The first two processes can not 
contribute to the inclusive spectra in the central region, but the third 
contribution is essential if the value of the intercept of the SJ exchange 
Regge-trajectory, $\alpha_{SJ}$, is large enough. At this point it is important 
to stress that since the quantum number content of the $SJ$ exchange matches 
that of the possible Odderon exchange, if the value of the $SJ$ 
Regge-trajectory intercept, $\alpha_{SJ}$, would turn out to be large and it 
would coincide with the value of the Odderon Regge-trajectory, 
$\alpha_{SJ}\simeq 0.9$, then the $SJ$ could be identified to the Odderon, or,
at last, to one baryonic Odderon component.

Let's finally note that the process shown in Fig.~14c can be very naturally 
realized in the quark combinatorial approach~\cite{AnSh} through the specific 
probabilities of a valence quark recombination (fusion) with sea quarks and 
antiquarks, the value of $\alpha_{SJ}$ depending on these specific 
probabilities.

The contribution of the graph in Fig.~14c is weighted in QGSM by coefficient 
$\varepsilon$ which determines the small probability for such a baryon 
number transfer to occur.

At high energies the SJ contribution to the inclusive cross section 
of secondary baryon production at large rapidity distance $\Delta y$ from 
the incident nucleon can be estimated as
\begin{equation}
(1/\sigma)d\sigma^B/dy \sim a_B\cdot\varepsilon\cdot
e^{(1 - \alpha_{SJ}) \Delta y} \;,
\end{equation}
where $a_B = a_N\cdot v^B_{ss}$. The baryon charge transferred to large 
rapidity distances can be determined by integration of Eq.~(35), so it is of 
the order of
\begin{equation}
\langle n_B \rangle \sim a_B\cdot\varepsilon /(1 - \alpha_{SJ}) \;.
\end{equation}
It is clear that the value $\alpha_{SJ} \geq 1$ should be excluded due to the 
violation of baryon-number conservation at asymptotically high energies.

\section{Comparison of the QGSM predictions with the experimental data}

Here we compare the results of QGSM predictions with all available experimental 
data on the $\bar{p}/p$ ratios at high energy.

To obtain the QGSM predictions for the $\bar{p}/p$ ratios we use the values 
of the probabilities $w_n$ in Eq.~(23) that are calculated in the frame of 
Reggeon theory \cite{KTM}, and the values of the normalization constants 
$a_{\pi}$ (pion production), $a_K$ (kaon production), $a_{\bar{N}}$ 
($B\bar{B}$ pair production), and $a_N$ (baryon production due to SJ 
diffusion) that were determined~\cite{KTM,KaPi,Sh} from the experimental 
data at fixed target energies. 

As an example of the experimental data description we present in Table 2 
\cite{AMPS} the calculated yields of different secondaries produced in $pp$ 
collisions at energy $\sqrt{s} = 200$ GeV in midrapidity region together with 
RHIC data (STAR Collaboration).

\begin{center}

\begin{tabular}{|c||r|r|r|} \hline
Particle & \multicolumn{3}{c|}{RHIC ($\sqrt{s} = 200$ GeV)}  \\ \cline{2-4}
& $\varepsilon = 0$ & $\varepsilon = 0.024$ & Experiment \cite{abe} \\  \hline
$\pi^+$         & 1.27    &        & $1.44 \pm 0.11$     \\ 
$\pi^-$         & 1.25    &        & $1.42 \pm 0.11$     \\
$K^+$           & 0.13    &        & $0.150 \pm 0.013$   \\
$K^-$           & 0.12    &        & $0.145 \pm 0.013$   \\
$p$             & 0.0755  & 0.0861 & $0.138 \pm 0.012$   \\
$\overline{p}$  & 0.0707  &        & $0.113 \pm 0.01$    \\
$\Lambda$       & 0.0328  & 0.0381 & $0.0385 \pm 0.0035$ \\
$\overline{\Lambda}$ & 0.0304  &   & $0.0351 \pm 0.0032$ \\
$\Xi^-$      & 0.00306  & 0.00359 & $0.0026 \pm 0.0009$   \\
$\overline{\Xi^+}$ & 0.00298 &     & $0.0029 \pm 0.001$   \\
$\Omega^-$       & 0.00020  & 0.00025 & *                 \\
$\overline{\Omega^+}$ & 0.00020  &    & *                 \\
\hline
\end{tabular}

$^*$ dn/dy($\Omega^- = \bar{\Omega}^+) = 0.00034 \pm 0.00019$

\end{center}

\noindent
Table 2. The QGSM results for midrapidity yields $dn/dy$ 
($\vert y \vert < 0.5$) for different secondaries at energy $\sqrt{s} = 200$ 
GeV. The results for $\varepsilon = 0.024$ are presented only when different 
from the case $\varepsilon = 0$. \\

In all cases our calculations generally underestimate the experimental 
points. However the theoretical calculations correspond to all inelastic $pp$ 
collisions, while the experimental data are obtained for events without 
single diffraction dissociation. In all cases, the agreement of the order of
10\% should be considered as good enough.

The ratio of $p$ to $\bar{p}$ yields at $y=0$ calculated with the QGSM is
shown in Fig.~15. The results with $\alpha_{SJ} = 0.9$ and 
$\varepsilon = 0.024$, $\alpha_{SJ} = 0.6$ and $\varepsilon = 0.057$, and 
$\alpha_{SJ} = 0.5$ and $\varepsilon = 0.0757$ are presented by dashed 
($\chi^2$/ndf=21.7/10), dotted ($\chi^2$/ndf=12.2/10), and dash-dotted 
($\chi^2$/ndf=11.1/10) curves, respectively. Thus, the most probable 
value of $\alpha_{SJ}$ from the point of view of the $\chi^2$ analysis is 
$\alpha_{SJ} = 0.5 \pm 0.1$.

\begin{figure}[htb]
\centering
\includegraphics[width=.55\hsize]{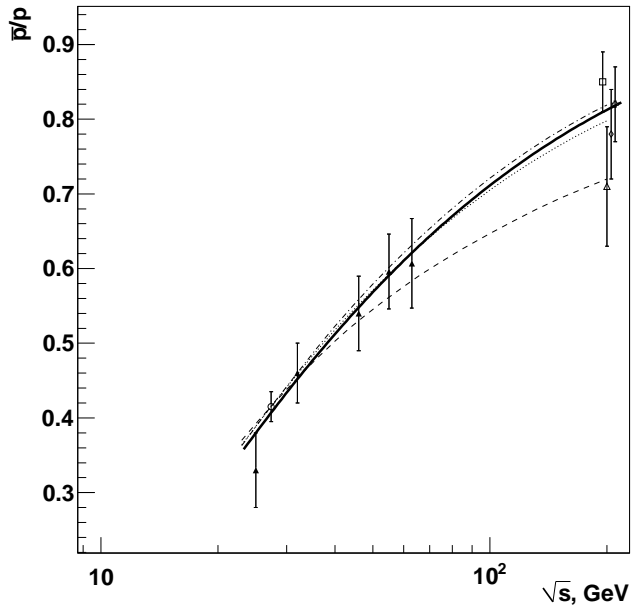}
\caption{\footnotesize 
The experimental ratios of $\bar{p}$ to $p$ production cross sections in 
high energies $pp$ collisions at $y=0$ \cite{Gue,Agu,BRA,PHO,PHE,STAR}, 
together with their fits \cite{MRS1} (solid curves), and 
by the QGSM description (dashed, dotted, and dash-dotted curves).}
\end{figure}

However, this conclusion comes from the global analysis of all experimental 
points, but one can see in Fig.~15 that the calculated value of $\bar{p}$ to 
$p$ production ratio with $\alpha_{SJ} = 0.9$ is in very good agreement with 
the experimental point of PHENIX Collaboration $0.71 \pm 0.01 \pm 0.08$
\cite{PHE}, so the situation is not clear.

\section{Predictions for $\bar{B}/B$ ratios at LHC}

We present in Tables 3 and 4 our predictiuons for antibaryon/baryon ratios 
in midrapidity region at energies $\sqrt{s} = 900$ GeV, 7 TeV, and 14 TeV 
for two possibilities of SJ contribution, $\alpha_{SJ} = 0.5$ 
($\omega$-reggeon contribution) and $\alpha_{SJ} = 0.9$ (Odderon 
contribution).

First of all we predict practically equal $\bar{B}/B$ ratios for the baryons 
with different strangeness, the small differences presented in Tables 3 and 4 
seem to be inside the accuracy of our calculations. The calculated 
$\bar{B}/B$ ratios do not practically depend either on the averaged transverse 
momenta of the considered secondaries.

At $\sqrt{s} = 900$ GeV we expect the values of $\bar{B}/B$ ratios to be 
about 0.96 in the case of $\alpha_{SJ} = 0.5$ ($\omega$-reggeon contribution) 
and about 0.90 in the case of $\alpha_{SJ} = 0.9$ (Odderon contribution). At 
$\sqrt{s} = 7$ TeV these ratios are predicted to be 0.99 and 0.95,
respectively. We do not present the predictions corresponding to no 
contribution of Reggeon with negative signature ($\varepsilon = 0$ in Eq.~(35)),
because it is in contradiction with the high energy data \cite{MRS,MRS1}.


\begin{center}

\vskip 5pt
\begin{tabular}{|c||r|r|r|r|} \hline
Ratio & \multicolumn{2}{c|}{$\sqrt{s} = 900$ GeV}  
& \multicolumn{2}{c|}{$\sqrt{s} = 7$ TeV}  \\ \cline{2-3} \cline{4-5} 
& $\alpha_{SJ} = 0.5$ & $\alpha_{SJ} = 0.9$  
& $\alpha_{SJ} = 0.5$ & $\alpha_{SJ} = 0.9$ \\   \hline
$\overline{p}/p$ & 0.955 & 0.892 & 0.989 & 0.946  \\
$\overline{\Lambda}/\Lambda$ & 0.949 & 0.887 & 0.986 & 0.945 \\
$\overline{\Xi}^+/\Xi^-$  & 0.965 & 0.909 & 0.991 & 0.958  \\
$\overline{\Omega}^+/\Omega^-$ & 0.965 & 0.907 & 0.992 & 0.958 \\
\hline
\end{tabular}

\end{center}

\noindent
Table 3. The QGSM predictions for antibaryon/baryon yields ratios in $pp$ 
collisions in midrapidity region ($\vert y \vert < 0.5$) for energies
$\sqrt{s} = 900$ GeV and $\sqrt{s} = 7$ TeV. Two possibilities are considered: 
$\alpha_{SJ} = 0.5$ ($\omega$-reggeon contribution) and $\alpha_{SJ} = 0.9$ 
(Odderon contribution).

At $\sqrt{s} = 14$ TeV the $\bar{B}/B$ ratios are predicted to be larger
than 0.99 for $\alpha_{SJ} = 0.5$ and about 0.96 for $\alpha_{SJ} = 0.9$.
So the experimental accuracy $\sim 1$ \% in these ratios is needed to  
discriminate between these two possibilities of $\alpha_{SJ}$ values.


\begin{center}

\begin{tabular}{|c||r|r|} \hline
Ratio & \multicolumn{2}{c|}{$\sqrt{s} = 14$ TeV} \\ \cline{2-3}
& $\alpha_{SJ} = 0.5$ & $\alpha_{SJ} = 0.9$ \\   \hline
$\overline{p}/p$ & 0.994 & 0.957  \\
$\overline{\Lambda}/\Lambda$  & 0.993 & 0.957 \\
$\overline{\Xi}^+/\Xi^-$   & 0.995 & 0.966  \\
$\overline{\Omega}^+/\Omega^-$ & 0.995 & 0.967  \\
\hline
\end{tabular}

\end{center}

\noindent
Table 4. The QGSM predictions for antibaryon/baryon yields ratios in $pp$ 
collisions in midrapidity region ($\vert y \vert < 0.5$) for $\sqrt{s} = 14$ 
TeV. Two possibilities are considered: $\alpha_{SJ} = 0.5$ ($\omega$-reggeon 
contribution) and $\alpha_{SJ} = 0.9$ (Odderon contribution).

The absolute values of midrapidity densities of produced secondaries are 
more model dependent in comparison with the antiparticle/particle ratios. We 
present in Table 5 the corresponding QGSM predictions at LHC energies 
$\sqrt{s} = 900$ GeV, $\sqrt{s} = 7$ TeV, and $\sqrt{s} = 14$ TeV. Baryon 
densities can be obtained with the help of Tables 3 and 4. 


The QGSM predictions for the spectra of secondary charged pions, charged 
kaons, protons, and antiprotons produced in $pp$ collisions at energies 
$\sqrt{s} = 900$ GeV and  $\sqrt{s} = 7$ TeV are presented in Fig.~16.

\begin{figure}[htb]
\centering
\includegraphics[width=.49\hsize]{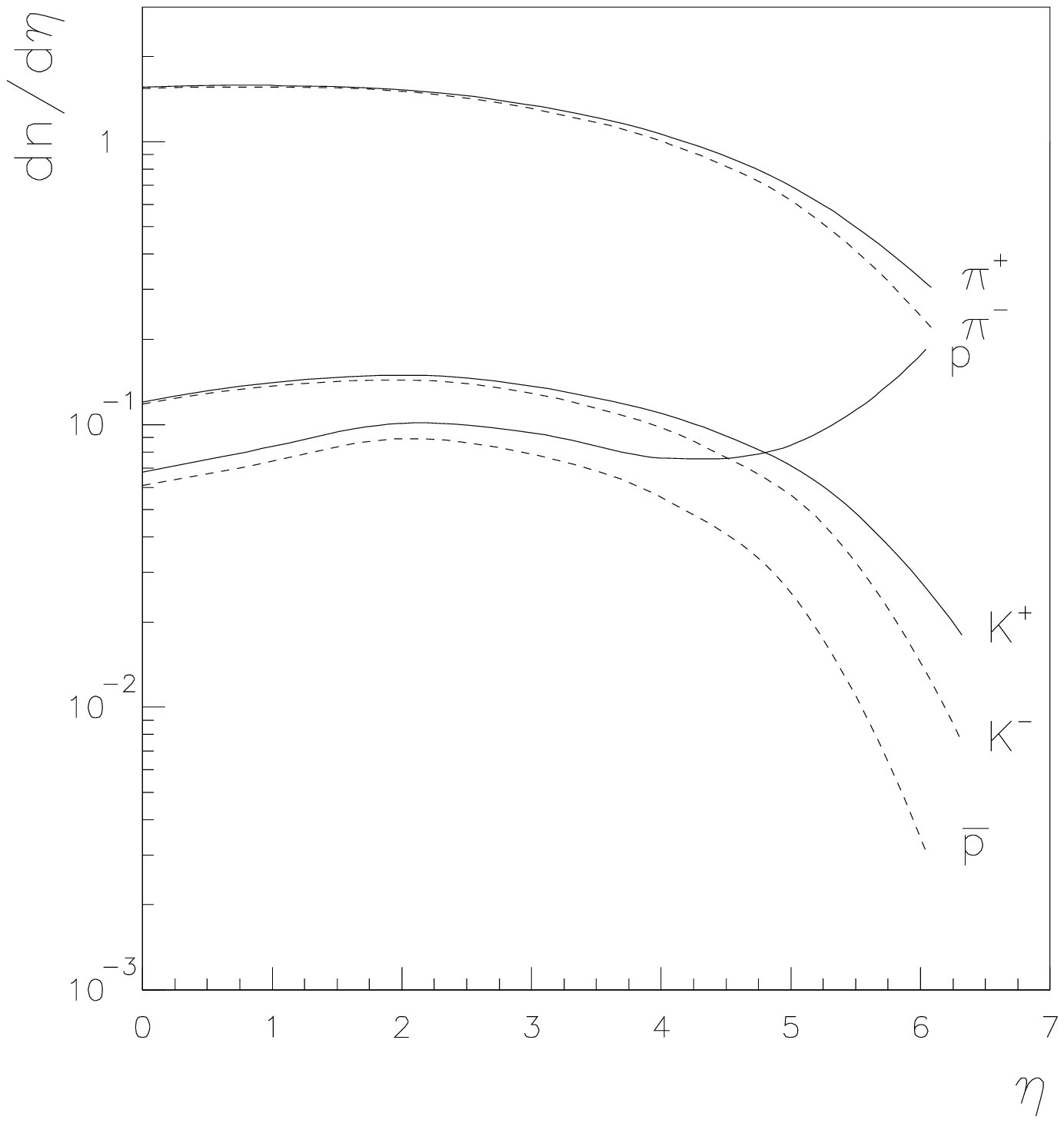}
\includegraphics[width=.49\hsize]{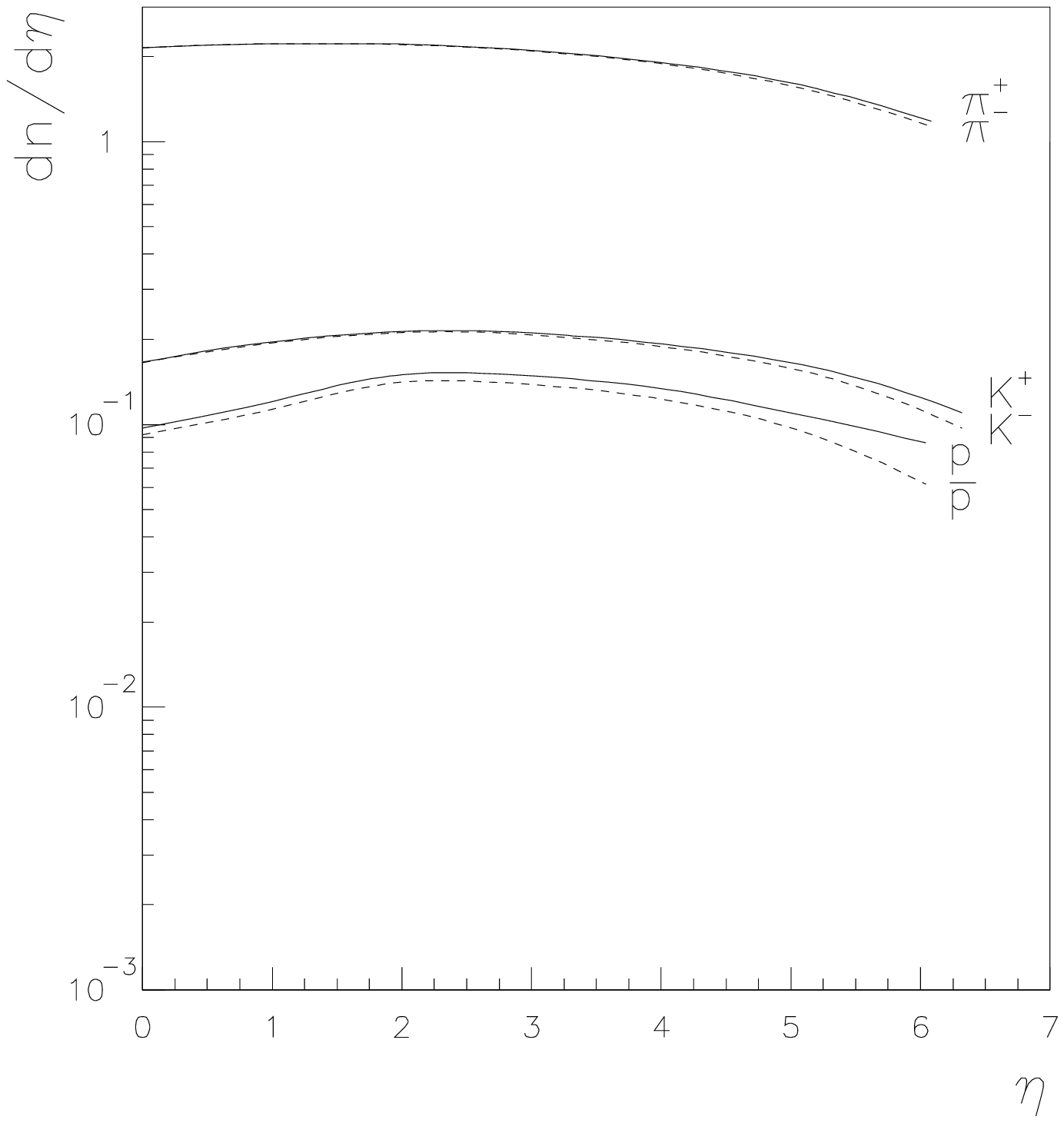}
\vskip -.6cm
\caption{\footnotesize 
The QGSM predictions for the spectra of secondary pions, kaons, protons and 
antiprotons at energies $\sqrt{s} = 900$ GeV (left panel) and $\sqrt{s} = 7$ TeV  (right panel).}
\end{figure}

\newpage

\begin{center}

\vskip 5pt

\begin{tabular}{|c||r|r|r|} \hline
Particle &  $\sqrt{s} = 900$ GeV & $\sqrt{s} = 7$ TeV
& $\sqrt{s} = 14$ TeV \\   \hline
$\pi^+$         & 1.68   & 2.32    & 2.54  \\ 
$\pi^-$         & 1.66   & 2.31    & 2.54   \\
$K^+$           & 0.17   & 0.23    & 0.25   \\
$K^-$           & 0.16   & 0.23    & 0.25   \\
$\overline{p}$  & 0.10  & 0.16   & 0.18      \\
$\overline{\Lambda}$  & 0.05   & 0.08   & 0.09   \\
$\overline{\Xi^+}$    & 0.005   & 0.009  & 0.011   \\
$\overline{\Omega^+}$ & 0.0004  & 0.0008 & 0.0009  \\
\hline
\end{tabular}

\end{center}

\noindent
Table 5. The QGSM results for the midrapidity yields $dn/dy$ 
($\vert y \vert < 0.5$) of different secondaries at LHC energies.

Preliminary data by the ALICE Collaboration \cite{ALI} presented in Table 6 
show that the $\omega$-reggeon contribution is enough for description of  
antiproton/proton ratios at LHC energies. If this is confirmed by further LHC data,
it would mean that the Odderon coupling must be very small.

\newpage

\begin{center}

\begin{tabular}{|c||r|r|} \hline
Variant  & $\sqrt{s} = 900$ GeV & $\sqrt{s} = 7$ TeV \\ \hline
$\alpha_{SJ} = 0.9$ & 0.89 & 0.95  \\
$\alpha_{SJ} = 0.5$   & 0.95 & 0.99   \\
Without               &     &    \\  
C-negative            & 0.98 & 1.   \\
exchange              &    &  \\ \hline
ALICE Coll. & $0.957 \pm $ & $0.991 \pm $ \\
            & $0.006 \pm 0.014$ & $0.005 \pm 0.014$ \\

\hline
\end{tabular}

\end{center}

\noindent
Table 6. The QGSM predictions for $\bar{p}/p$ in $pp$ collisions at LHC 
energies and the data by the ALICE Collaboration \cite{ALI}. 

The LHCb Collaboration plans to measure the ratios of antibaryons 
to baryons spectra in some interval of pseudorapidities. 
Thinking of this, we present in Figs.~17 and 18 the $\eta$-dependences of
$\bar{p}/p$ and $\bar{\Lambda}/\Lambda$ at energies 
$\sqrt{s} = 900$ GeV and $\sqrt{s} = 7$ TeV. 

\begin{figure}[htb]
\centering
\vskip .5cm
\includegraphics[width=.49\hsize]{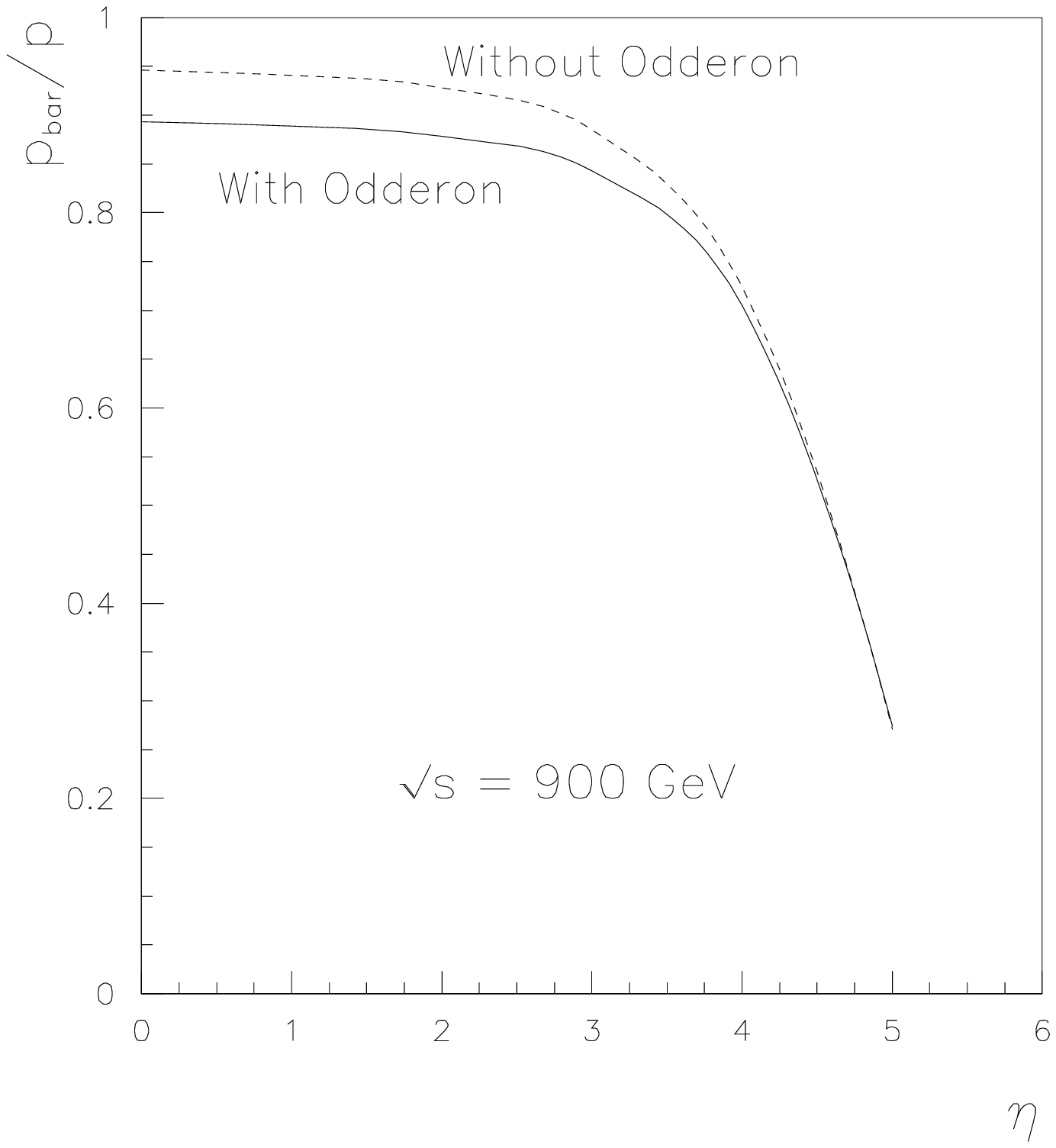}
\includegraphics[width=.49\hsize]{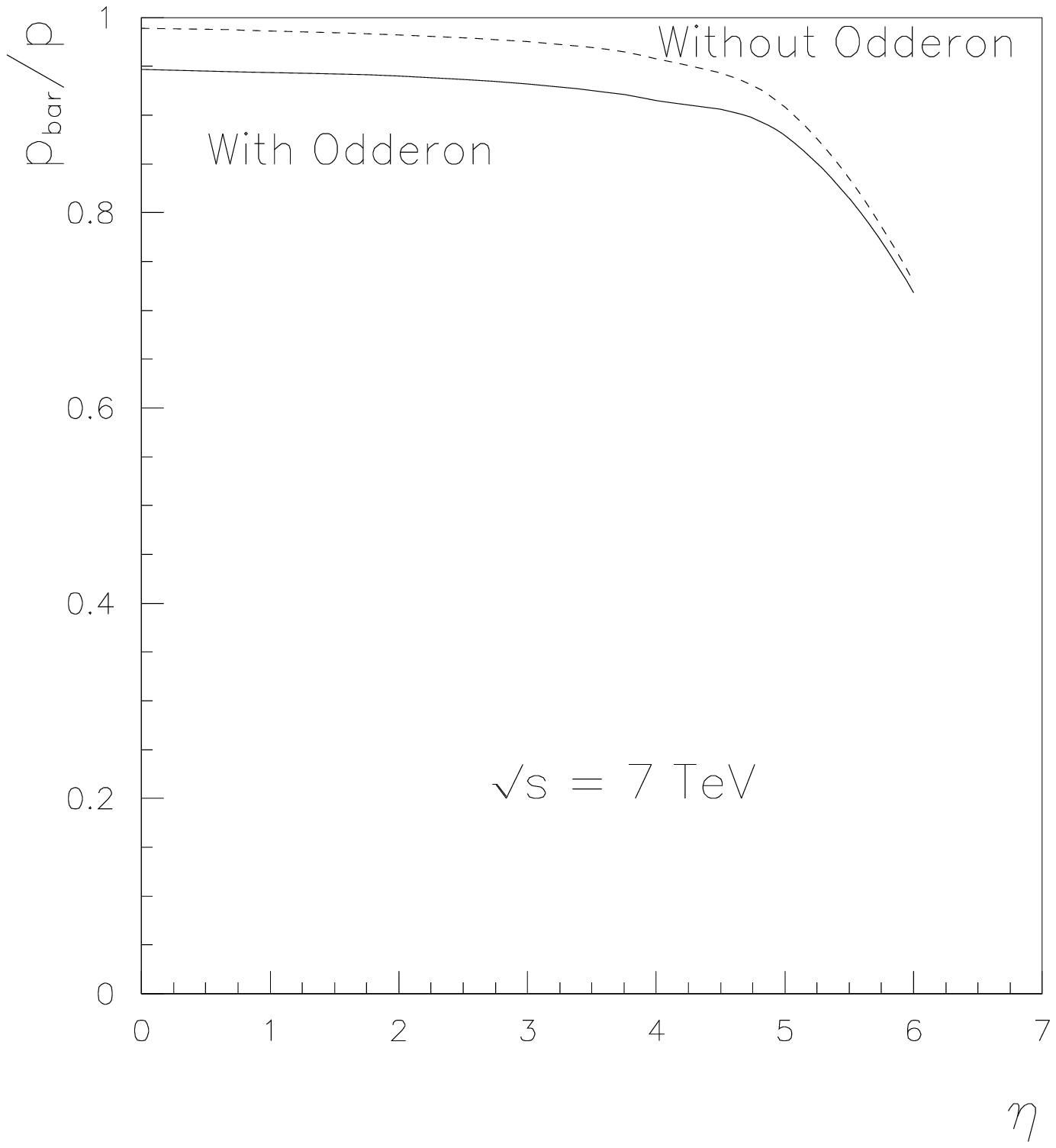}
\caption{\footnotesize 
The QGSM predictions for the ratios of the spectra of secondary antiprotons 
to protons as the functions of their pseudorapitities at energies 
$\sqrt{s} = 900$ GeV (left panel) and $\sqrt{s} = 7$ TeV  (right panel).}
\end{figure}

\newpage

\begin{figure}[htb]
\centering
\vskip .5cm
\includegraphics[width=.49\hsize]{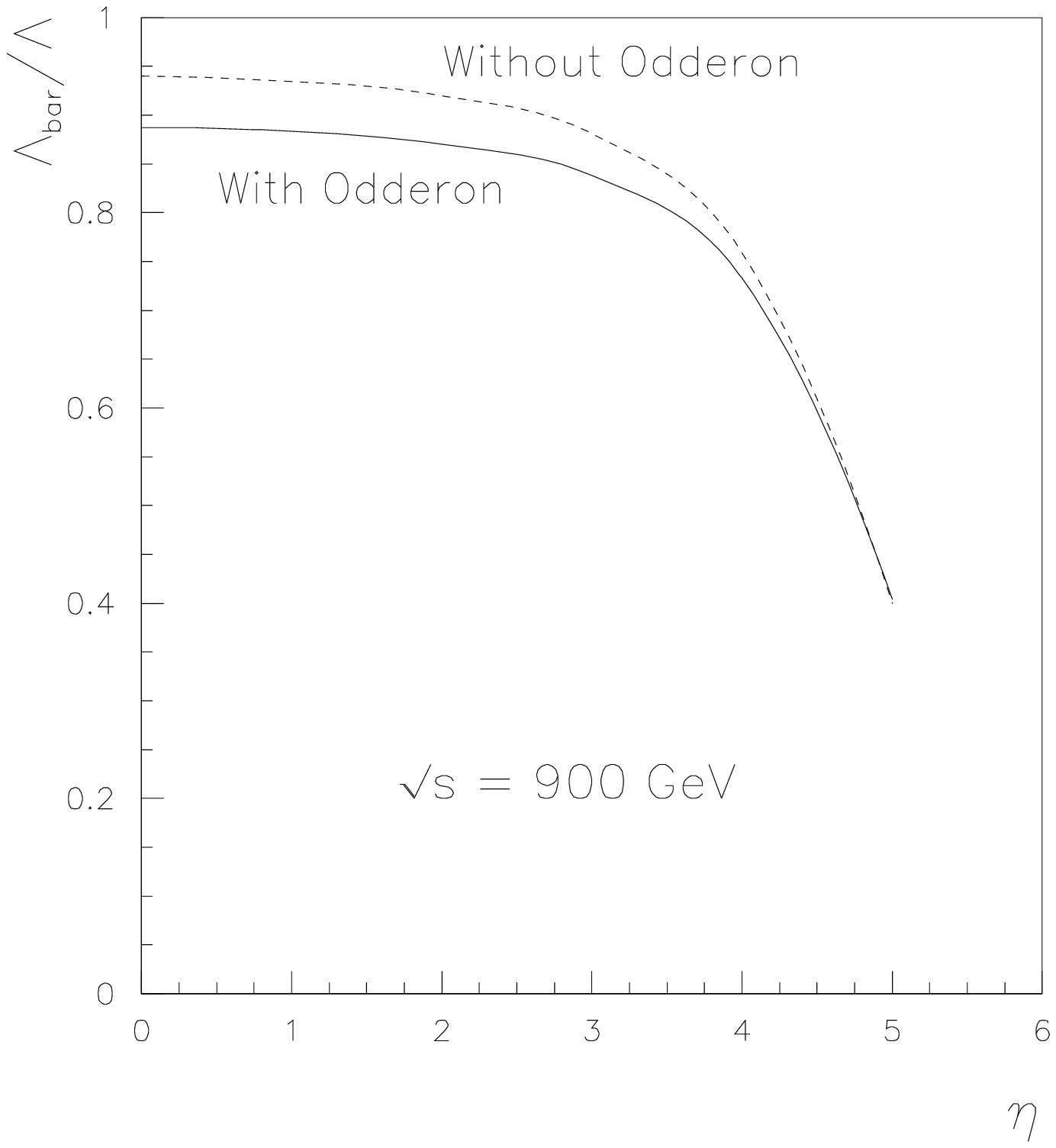}
\includegraphics[width=.49\hsize]{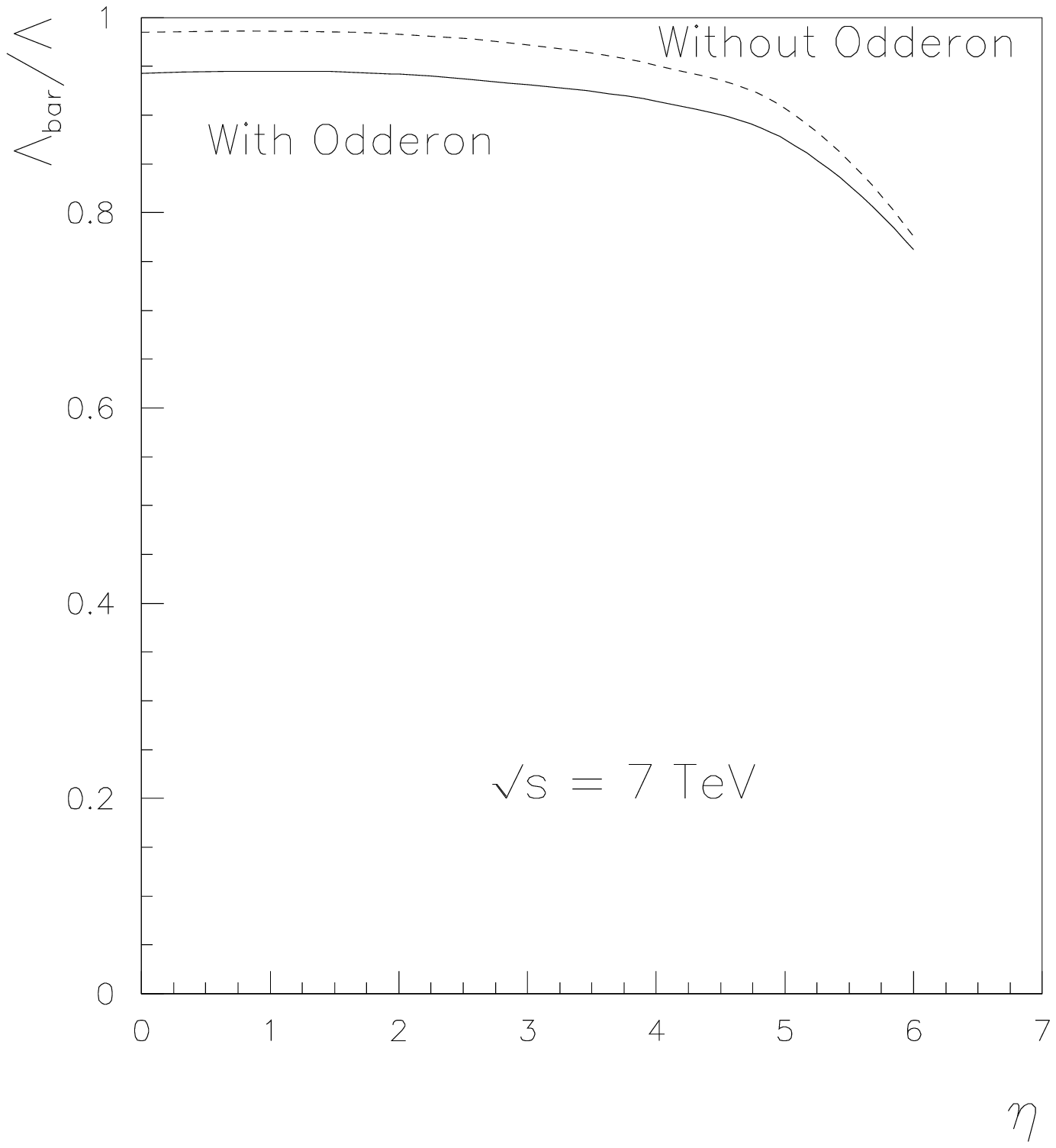}
\caption{\footnotesize 
The QGSM predictions for the ratios of the spectra of secondary 
$\bar{\Lambda}$ to $\Lambda$ as the functions of their pseudorapitities 
at energies $\sqrt{s} = 900$ GeV (left panel) and $\sqrt{s} = 7$ TeV  
(right panel).}
\end{figure}


\section{Conclusion}

The first experimental data obtained at LHC are in general agreement with
the calculations provided in the framework of Regge theory and of the 
QGSM with the same values of parameters which were determined at lower 
energies (mainly for description the fixed target experiments). 

We neglect the possibility of interactions between Pomerons (so-called 
enhancement diagrams) in the calculations of integrated cross sections and
inclusive densities. Such interactions are very important in the cases of 
heavy ion \cite{CKT} and nucleon-nucleus \cite{MPS} interactions at RHIC 
energies, and their contribution should increase with energy. However we 
estimate that the contributions of these enhanced diagrams inclusive density 
of secondaries produced in $pp$ collisions at LHC energies is not large enough 
to be significant.

In the case of the inclusive production of particles and antiparticles in 
central (midrapidity) region in $pp$ collisions the only evidence for the 
Odderon exchange with $\alpha_{Odd}(0) \simeq 0.9$ in the inclusive 
reactions is proveded by two experimental points for $\bar{B}B$ production 
asymmetry obtained by the H1 Collaboration~\cite{H1,H1a}. The first point 
\cite{H1} (for $\bar{p}/p$ ratio) is until now not published, and the second 
one \cite{H1a} (for $\bar{\Lambda}/\Lambda$ ratio) shows a very large error bar.
On the other hand, only for these two points the kinematics would allow 
the energy of the Odderon exchange to be large enough 
($\sqrt{s} \simeq 10^2$ GeV) to be noticed.

ALICE Collaboration data are in disagreement with a numerically large 
contribution of the Odderon with $\alpha_{SJ} = 0.9$, the coupling of such
an Odderon should be suppressed.

ALICE Collaboration data are in agreement with the only $\omega$-Reggeon contribution.

One has to expect that furhter LHC data will make the situation more clear. 

{\bf Acknowledgements}

We are grateful to Ya.I. Azimov, A.B. Kaidalov, and M.G. Ryskin for useful 
discussions and comments. This paper was supported by Ministerio de Educaci\'on 
y Ciencia of Spain under the Spanish Consolider-Ingenio 2010 Programme CPAN 
(CSD2007-00042) by Xunta de Galicia project FPA 2005--01963 and, in part, 
by grant RSGSS-3628.2008.2.


\end{document}